\documentclass[twocolumn]{aastex631}
\usepackage[utf8]{inputenc}

\usepackage{xcolor,tikz}
\usepackage{gensymb}
\usepackage{multirow}

\accepted{\today}

\usepackage{hyperref}

\begin{document}



\title{Multiperiodic pulsations of the unique DAQ white dwarf J0551$+$4135: insights into a merger remnant}

\author[0000-0002-6603-994X]{Murat Uzundag}
\affiliation{Institute of Astronomy, KU Leuven, Celestijnenlaan 200D, 3001, Leuven, Belgium}

\author[0000-0001-6098-2235]{Mukremin Kilic} 
\affiliation{Homer L. Dodge Department of Physics and Astronomy, University of Oklahoma, 440 W. Brooks St., Norman, OK, 73019, USA}
    
\author[0000-0002-8079-0772]{Francisco C. De Gerónimo}
\affiliation{Grupo de Evolución Estelar y Pulsaciones. Facultad de Ciencias Astronómicas y Geofísicas, Universidad Nacional de La Plata, Paseo del
Bosque s/n, (1900) La Plata, Argentina}
\affiliation{Instituto de Astrofísica de La Plata, IALP (CCT La Plata), CONICET-UNLP, Argentina}

\author[0000-0002-0006-9900]{Alejandro H. Córsico}
\affiliation{Grupo de Evolución Estelar y Pulsaciones. Facultad de Ciencias Astronómicas y Geofísicas, Universidad Nacional de La Plata, Paseo del
Bosque s/n, (1900) La Plata, Argentina}
\affiliation{Instituto de Astrofísica de La Plata, IALP (CCT La Plata), CONICET-UNLP, Argentina}

\author[0009-0009-9105-7865]{Gracyn Jewett} 
\affiliation{Homer L. Dodge Department of Physics and Astronomy, University of Oklahoma, 440 W. Brooks St., Norman, OK, 73019, USA}

\author[0000-0001-7143-0890]{Adam Moss} 
\affiliation{Department of Astronomy, University of Florida, Bryant Space Science Center, Stadium Road, Gainesville, FL 32611, USA}

\author[0000-0002-6153-7173]{Alberto Rebassa-Mansergas}
\affiliation{Departament de Física, Universitat Politècnica de Catalunya, c/Esteve Terrades 5, 08860 Castelldefels, Spain}
\affiliation{Institut d'Estudis Espacials de Catalunya, Esteve Terradas, 1, Edifici RDIT, Campus PMT-UPC, 08860 Castelldefels, Barcelona, Spain}

\author[0000-0002-3316-7240]{Alex J. Brown}
\affiliation{Hamburger Sternwarte, University of Hamburg, Gojenbergsweg 112, 21029 Hamburg, Germany}

\author[0000-0002-3524-190X]{Maria  Camisassa}
\affiliation{Departament de Física, Universitat Politècnica de Catalunya, c/Esteve Terrades 5, 08860 Castelldefels, Spain}

\author[0000-0003-2368-345X]{Pierre Bergeron} 
\affiliation{Département de Physique, Université de Montréal, C.P. 6128, Succ. Centre-Ville, Montréal, QC H3C 3J7, Canada}




\begin{abstract}

2MASS J05513444+4135297 (herafter J0551+4135) is the only pulsating DAQ white dwarf known with a carbon and hydrogen atmosphere. Its unusual atmospheric composition and kinematics indicate a white dwarf merger origin. We present time-series photometry of J0551+4135 obtained using the Apache Point Observatory 3.5m, Gemini North 8m, and Gran Telescopio Canarias 10m telescopes. 
 J0551+4135 exhibits variations in pulsation amplitude and frequency over time. We detect ten significant recurring peaks across different subsets of observations, with frequencies ranging from 987 to 1180~$\mu$Hz, consistent with non-radial gravity ($g$)-mode oscillations.
We present new evolutionary models suitable for spectroscopic characterization of DAQ white dwarfs, and derive a mass of
$1.13 \pm 0.01\,M_\odot$ and a cooling age of $1.7 \pm 0.1$ Gyr for a CO core, and $1.12 \pm 0.01\,M_\odot$ and
$1.6 \pm 0.1$\,Gyr for an ONe-core white dwarf, respectively. However, detailed asteroseismology of this unique pulsator
has to wait until fully-consistent DAQ evolutionary models are available. Further observations, including multi-site campaigns to reduce daily aliasing and to improve the signal-to-noise ratio would be helpful for identification of additional modes and constraining the internal structure of this unique pulsator.
\end{abstract}

\section{Introduction} 
\label{intro}

White dwarfs (WDs) are the final evolutionary stage of most stars —specifically, low- and intermediate-mass stars that make up over 95\% of the Milky Way’s stellar population \citep{2010A&ARv..18..471A,2022PhR...988....1S}. As they cool, many WDs pass through a phase of pulsational instability, becoming variable stars. Pulsating Hydrogen-rich (DAV) WDs or ZZ Ceti stars show non-radial gravity($g$)-mode pulsations with periods between $\sim$100 and 1400 s, driven within a narrow effective temperature range of $\sim10,500$ – $13,500$ K \citep{Fontaine08, Winget08, 2019A&ARv..27....7C}. 
Asteroseismology uses the observed pulsation periods to study the interiors of these dense stellar remnants \citep[e.g.,][]{1998ApJS..116..307B, 2005ApJ...622..572B, 2019A&A...632A.119C, 2019A&A...621A.100D, 2023MNRAS.522.2181K, 2025ApJ...988...32C}.

The mass distribution of pulsating DA WDs is strongly peaked around $0.6\,M_\odot$, consistent with expectations from single-star evolution \citep{2010A&ARv..18..471A}. Indeed, most WDs are found within the $0.45$–$1.05\,M_\odot$ range, and are believed to possess carbon–oxygen (CO) cores, the typical end-products of He-burning in low- and intermediate-mass progenitors. Below $0.45\,M_\odot$, white dwarfs are expected to harbor He cores, often resulting from binary evolution that prevents the progenitor from reaching the He-burning phase \citep{2004ApJ...612L..25N, Rebassa2011, 2013A&A...557A..19A}.

At the high-mass end, the internal composition of ultra-massive ($M_{\rm WD} \gtrsim 1.05\,M_\odot$) WDs remains a subject of active debate. These stars may contain ONe cores formed via off-center carbon ignition in more massive progenitors \citep{2010A&A...512A..10S}, or even CO cores formed through scenarios which involve the rotation of the degenerate core after core helium burning and reduced mass-loss rates in massive asymptotic giant-branch stars \citep{2021A&A...646A..30A}. Alternatively, they may possess hybrid CO–Ne cores if carbon burning was incomplete or only partially developed \citep{2024ApJ...975..259D, 2022A&A...659A.150D, 2013ApJ...772...37D}. The origin of ultra-massive WDs also remains elusive— a significant fraction may form through mergers \citep{temmink20,kilic23,jewett24}, which could lead to CO core merger remnants. 

Most pulsating DA white dwarfs show H-dominated spectra. In rare cases, heavier elements, likely from planetary debris accretion, appear, as seen in G29-38, suggesting complex evolutionary histories and external pollution \citep[e.g.,][]{2023MNRAS.526.2846U}. A notable discovery by \citet[][see also \citealt{vincent20}]{2020NatAs...4..663H} identified a new subclass of pulsators: DAQ white dwarfs, which have hydrogen atmospheres with a substantial amount of carbon. 
The unique hydrogen-carbon atmospheres, kinematics, and large masses of DAQ white dwarfs suggest they arise from unusual evolutionary paths, such as white dwarf mergers \citep{dunlap15,coutu19, kawka23,2020NatAs...4..663H, kilic24,kilic25b}.

WD~J0551+4135 stands out as a particularly intriguing member of the DAQ class. With a mass of $M = 1.13\,M_\odot$ and a mixed H/C atmosphere (C/H $\sim$ 0.15 by number), it challenges standard single-star evolutionary predictions. Detailed spectroscopic modeling reveals that both H and He layers must be extremely thin ($M_{\rm H} < 10^{-9.5}\,M_\star$, $M_{\rm He} < 10^{-7.0}\,M_\star$), suggesting a highly stripped envelope \citep{2020NatAs...4..663H}. Combined with its high velocity with respect to the local standard of rest ($129 \pm 5$ km s$^{-1}$), this peculiar envelope composition supports a formation channel involving the merger of two white dwarfs in a compact binary system.

The unique spectral and structural properties of WD~J0551+4135 make it an exceptional laboratory to explore the aftermath of WD mergers. \citet[][see also \citealt{vincent20}]{2020NatAs...4..663H} reported low-amplitude (0.4–0.6\%) photometric variations at a single period of 840 s. 
Here, we present extended follow-up photometric observations that significantly enhance the detection of pulsation modes in this star. Our observational campaign is detailed in Section~\ref{obs}, followed by the presentation of light curves and frequency analysis in Section~\ref{analysis}. 
The derivation of the stellar mass is discussed in 
Section~\ref{evolutionary_models}, and we conclude with a summary of our findings in Section~\ref{conclusions}.

\section{Photometric observations }
\label{obs}

We acquired high-speed photometry of J0551+4135 on 12 nights between UT 2023 Nov 24 and 2025 Apr 12 at the Apache Point
Observatory (APO) 3.5 m telescope. We used the Astrophysical Research Consortium Telescope Imaging Camera (ARCTIC) and the BG40 filter.
We obtained 8-10 s long back-to-back exposures on each night. To reduce the read-out time, we used the quad amplifier mode and binned the
CCD by $3\times3$, which resulted in a plate scale of 0.342 arcsec pixel$^{-1}$ and a read out time of 4.5 s. This resulted in a cadence of
12.5 to 14.5 s on each night. In total, we obtained 34.8 h of observations at the APO 3.5m telescope. 

\begin{table}
\setlength{\tabcolsep}{3pt}
\centering
\caption{Observation log for WD~J0551$+$4135.}
\begin{tabular}{cccc}
\hline
\hline
UT Date & Instrument & Length of Obs. & Exp. time \\
(yyyy-mm-dd) &  & (h) & (s) \\
\hline
2023-11-24 & ARCTIC & 2.7 & 10 \\
2023-12-30 & ARCTIC & 3.6 & 10 \\
2024-01-07 & ARCTIC & 2.0 & 10 \\
2024-01-19 & ARCTIC & 4.8 & 10 \\
2024-10-09 & ARCTIC & 2.7 & 10 \\
2024-10-14 & GMOS & 2.0 & 7 \\
2024-11-22 & HIPERCAM & 1.5 & 2.1/0.4/0.8/0.8/1.2 \\
2024-11-23 & ARCTIC & 2.9 & 8 \\
2024-11-29 & GMOS & 2.0 & 7 \\
2024-12-05 & GMOS & 2.8 & 7 \\
2024-12-20 & GMOS & 2.4 & 7 \\
2024-12-21 & ARCTIC & 4.0 & 10 \\
2024-12-24 & GMOS & 1.9 & 7\\
2024-12-27 & GMOS & 2.0 & 7 \\
2025-01-08 & HIPERCAM & 4.0 & 2.1/0.4/0.8/0.8/1.2 \\
2025-02-03 & ARCTIC & 3.0 & 10\\
2025-02-05 & ARCTIC & 2.0 & 10 \\
2025-02-09 & ARCTIC & 2.2 & 8 \\
2025-03-09 & ARCTIC & 2.5 & 10 \\
2025-04-12 & ARCTIC & 2.4 & 10 \\
\hline
\end{tabular}
\label{tab:ObsLogs}
\end{table}

We obtained additional time series photometry of J0551+4135 using the Gemini North 8m telescope on six nights between UT 2024 Oct 14 and
Dec 27. We used the Gemini Multi Object Spectrograph (GMOS) in the imaging mode with the $g$ filter as part of the queue programs
GN-2024B-Q-205 and GN-2024B-Q-303. We obtained 293, 293, 292, 190, 287, and 287 back-to-back exposures over those six nights, respectively.
To reduce the read out time, we binned the chip by $4\times4$, which resulted in a plate scale of 0.32 arcsec pixel$^{-1}$ and a 17.5 s
overhead, resulting in a cadence of 24.5 s with our 7 s long exposures. The observations obtained on UT 2024 Dec 20 suffered from
significant delays in writing the images to the disk for an unknown reason, lowering our cadence. These data are excluded from our
frequency analysis. 

We were able to obtain additional observations of J0551+4135 using the HiPERCAM multi-band imager \citep{dhillon21} on the 10.4 m Gran Telescopio Canarias (GTC) on UT 2024 Nov 22 and 2025 Jan 08.
We obtained continuous observations of 1.5~h and 4.0~h on the two nights, respectively, with exposure times of 2.1/0.4/0.8/0.8/1.2~s in the Super–Sloan Digital Sky Survey (SDSS) $u_{s}/g_{s}/r_{s}/i_{s}/z_{s}$ filters simultaneously and with the detectors binned $4\times4$.
HiPERCAM uses frame-transfer CCDs so dead time between exposures is negligible.
We used the HiPERCAM data reduction pipeline \citep{dhillon21} to debias, flat-field, and fringe subtract the frames before extracting differential photometry relative to a bright
comparison star in the field. We used a variable target aperture size, set to $1.0 \times FWHM$, to minimize variable seeing effects and
to maximize the signal-to-noise ratio (S/N) of the extracted photometry.

\section{Results}
\label{analysis}

\subsection{Light curve analysis}
\label{LC_analysis}

\begin{figure*}
\includegraphics[width=1.0\textwidth]{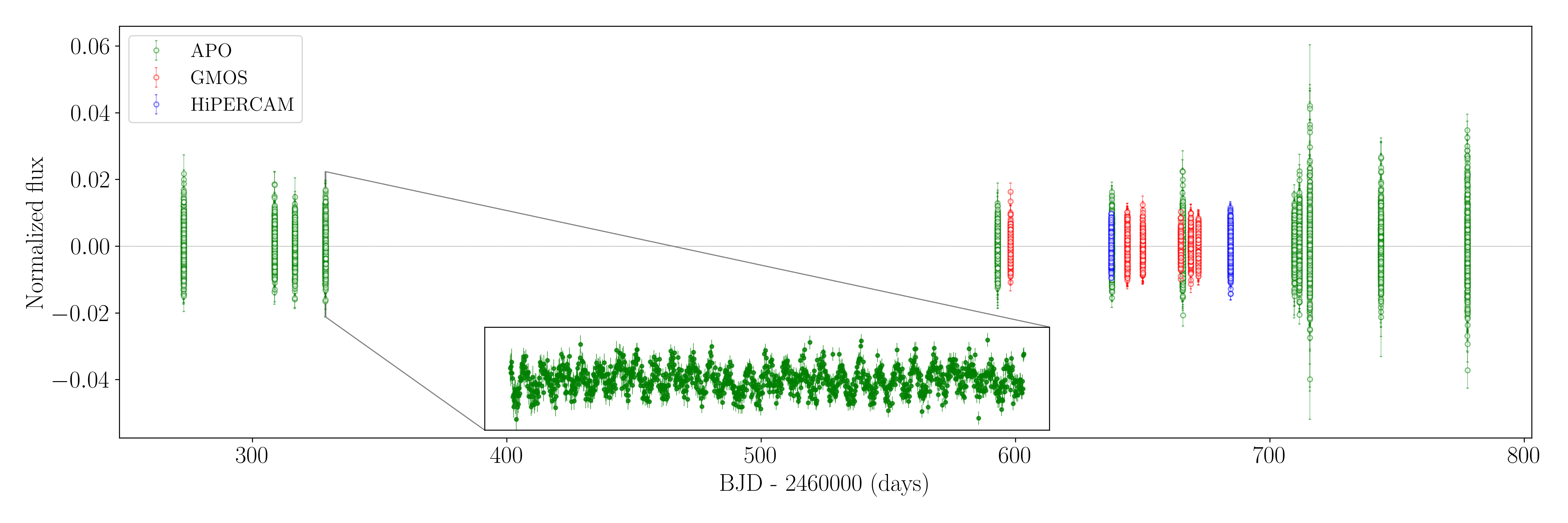} 
\caption{Combined normalized light curve of J0551 obtained with ARCTIC (green), GMOS (red), and HiPERCAM (blue) between November 2023 and April 2025. Each point represents normalized flux as a function of Barycentric Julian Date (BJD – 2460000). The data from different instruments were normalized by subtracting either a mean or unity flux level to ensure consistency across observing runs. The dashed horizontal line marks the zero-flux reference. The inset shows a zoomed-in view of the ARCTIC observations from 2024 January 19, covering a 4.8-hour sequence that reveals short-term variability on timescales of minutes. See section \protect\ref{analysis} for more details. }
\label{fig:LC_full}
\end{figure*}

Time-series photometry of J0551+4135 was obtained with APO, Gemini, and the GTC. Figure~\ref{fig:LC_full} shows the combined light curve from all three telescopes, and Table~\ref{tab:ObsLogs} summarizes the observations used in this study. We analyzed these light curves using a two-step procedure designed to optimize the reliability of the frequency analysis \citep[see also][]{2025ApJ...980L...9D, 2025ApJ...988...32C}. First, we applied a $5\sigma$ clipping algorithm to remove outliers from the raw flux data. We then detrended the resulting light curves by fitting and subtracting a second-order polynomial to eliminate long-term variations, such as residual atmospheric or instrumental trends, that could obscure low-amplitude periodicities. This approach effectively isolates the short-term stellar variability while preserving the intrinsic signal.

\subsection{Frequency Analysis} \label{sec:freq_analysis}

To investigate the pulsational behavior of our target, we performed a detailed frequency analysis using all available light curves obtained between 2019 \citep[from][]{2020NatAs...4..663H} and 2025 (this work). This extensive, multi-year dataset allows a characterization of the star’s pulsation spectrum and enables us to distinguish persistent pulsation modes from transient or instrumental features.

\begin{figure*}
   \includegraphics[width=1.0\textwidth]{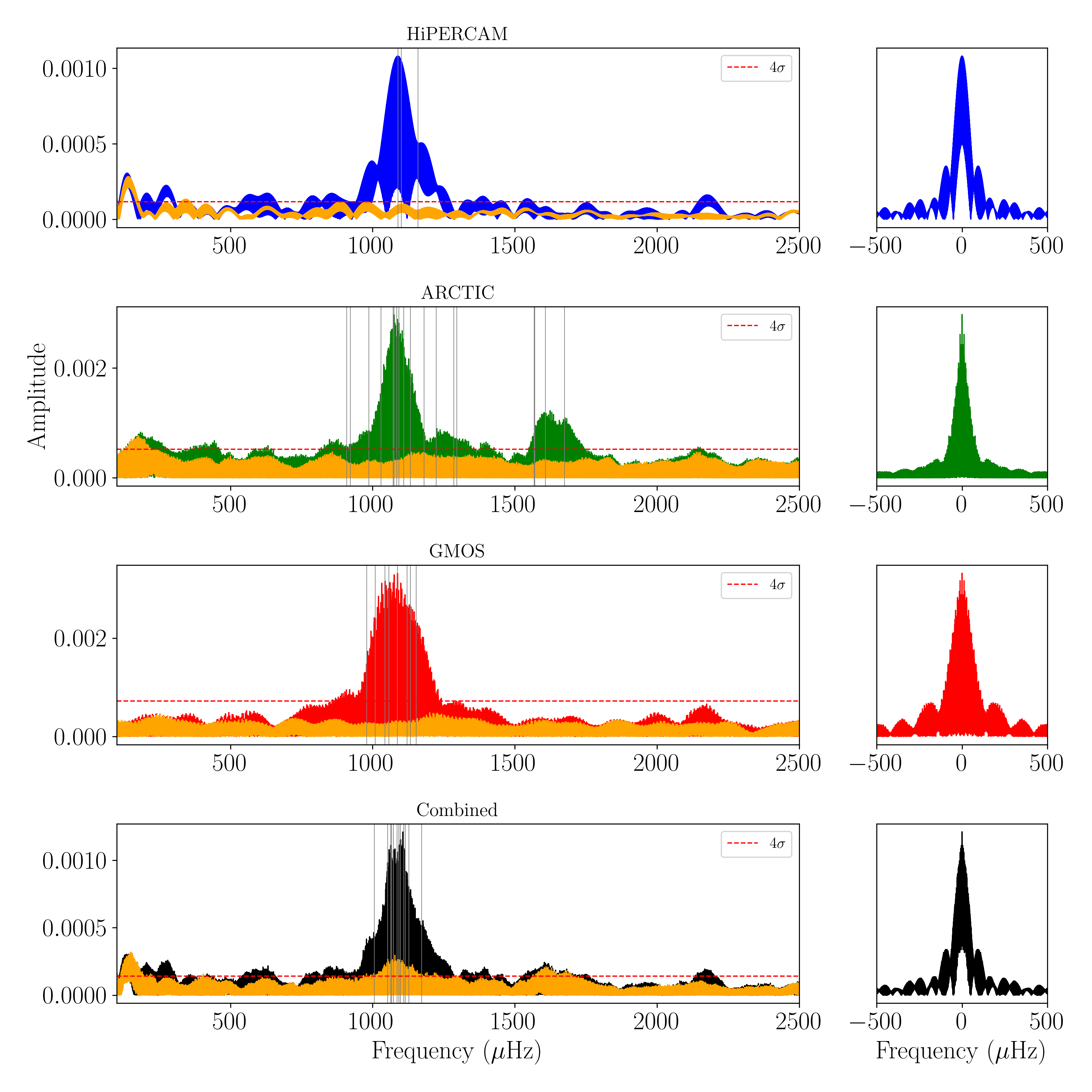} 
\caption{Lomb–Scargle periodograms (left panel) and window functions (right panel) for J0551 from different datasets: HiPERCAM (blue), ARCTIC (green), GMOS (red), and the combined dataset (black). The horizontal dashed red line indicates the 4$\sigma$ noise threshold computed from the mean noise level. The orange curves show the residual (prewhitened) periodograms after removing the identified frequencies. Vertical grey lines mark the prewhitened frequencies in each dataset.}
\label{fig:FT_full}
\end{figure*}

Figure~\ref{fig:FT_full} presents the Lomb--Scargle periodograms and the corresponding window functions for each instrument, as well as for the combined dataset. Each row corresponds to one instrument: HiPERCAM (blue), ARCTIC (green), GMOS (red), and the combined light curve (black). The left panels display the amplitude spectra as a function of frequency, while the right panels show the window functions, which illustrate the spectral resolution and aliasing patterns characteristic of each dataset. For each periodogram, we computed the median noise level and adopted a detection threshold corresponding to a signal-to-noise ratio (S/N) of 4, following the criterion of \citet{1993A&A...271..482B} and \citet{2008A&A...477..627C}. This significance threshold is indicated by red dashed lines in Figure~\ref{fig:FT_full}. The most prominent pulsation peaks cluster between 1000 and 1200~$\mu$Hz, consistently present in all subsets and in the combined periodogram. A secondary group of peaks appears between 1595 and 1700~$\mu$Hz, detected primarily in four ARCTIC runs obtained between 2023 November 24 and 2024 January 19.
In Appendix~\ref{LC}, we present a detailed analysis of each individual observing campaign, including the nightly light curves and their corresponding Fourier transforms.

In the combined dataset, resolving individual pulsation peaks is challenging due to the large temporal gaps between observing campaigns, which introduce strong aliasing (see the right panel of Fig.~\ref{fig:FT_full}). The window functions display complex structures with daily side-lobes and multi-month gaps, reflecting the discontinuous temporal coverage and varying spectral resolutions of the instruments. These effects, together with bandpass-dependent amplitude variations, complicate the prewhitening process: amplitude and frequency variability in several modes produced residual excess power after subtraction, and closely spaced peaks sometimes required fitting only the dominant component to avoid spurious detections. Our photometric monitoring spans from 2023 November 24 to 2025 April 12, combining data from ARCTIC, GMOS, and HiPERCAM. The full baseline of $\Delta T = 505$~days yields a formal Rayleigh frequency resolution of $\Delta f = 1/\Delta T = 0.023~\mu$Hz. However, the large gaps between campaigns significantly degrade the effective resolving power. To quantify this, we applied the criterion of \citet{LD1978}, $\Delta f = 1.5/\Delta T$. Including the gaps, the effective resolution is $\Delta f_{\mathrm{full}} = 0.034~\mu$Hz, while for the individual campaigns (excluding gaps) we find $\Delta f_{\mathrm{ARCTIC}} = 0.31~\mu$Hz, $\Delta f_{\mathrm{GMOS}} = 0.41~\mu$Hz, and $\Delta f_{\mathrm{HiPERCAM}} = 1.47~\mu$Hz, corresponding to temporal coverages of 56, 42, and 12~days, respectively. The spectral window function of the combined dataset exhibits strong daily aliases at $\simeq11.57~\mu$Hz and secondary structures caused by the long temporal gaps. These aliases were carefully considered during the prewhitening and frequency selection process. For the individual subsets, the achievable frequency resolution ranges from $\sim0.12$ to $0.21~\mu$Hz, setting the practical limit for resolving and identifying independent pulsation modes within each campaign.

We carried out an iterative prewhitening analysis using both the combined light curve and each individual subset. Figure~\ref{fig:FT_full} presents the Lomb--Scargle periodograms (left panels) and window functions (right panels) for HiPERCAM (blue), ARCTIC (green), GMOS (red), and the combined dataset (black). The fluxes were median-subtracted, and a uniform frequency grid between 0 and 3470~$\mu$Hz (corresponding to 0--300~d$^{-1}$) was computed with an oversampling factor of 10. For each iteration, we calculated a Lomb--Scargle periodogram \citep{1976Ap&SS..39..447L,1982ApJ...263..835S}, highlighting both the original amplitude spectrum and the residuals after prewhitening. The horizontal dashed red lines indicate the 4$\sigma$ noise threshold computed from the median noise level, while the orange curves show the residual periodograms after removing the identified frequencies. Vertical grey lines mark the prewhitened frequencies in each dataset. The noise level was defined as the median of the amplitude spectrum, and significant frequencies were required to satisfy $\mathrm{S/N} \ge 4$. Diagnostic plots and residual light curves from each iteration were inspected to verify the extracted frequencies. All significant detections are listed in Table~\ref{tab:Frequencies}. 

Across all datasets from 2019 to 2025, including APO, Gemini, and HiPERCAM observations, we find a consistent set of candidates of pulsation modes. Although many peaks exceed the 4$\times$S/N detection limit in individual runs, several correspond to the same underlying frequencies distorted by daily aliasing or limited sampling. All prewhitened peaks from combined dataset and each subset are reported in Table~\ref{tab:Frequencies}.
We also search for possible combination frequencies computing all linear sums and differences between the detected modes following the method of \citet{2023MNRAS.526.2846U}. All combination candidates for each subset beyond 2100~$\mu$Hz are reported in Table~\ref{tab:Frequencies}, though the current dataset is insufficient to confirm them conclusively. The limited baseline and frequency resolution hinder the detection of low-amplitude or closely spaced combinations, especially at higher frequencies. Extended, continuous photometric monitoring will be required to resolve these ambiguities and confirm potential combination signals.

\begin{figure*}
\centering
   \includegraphics[width=1.0\textwidth]{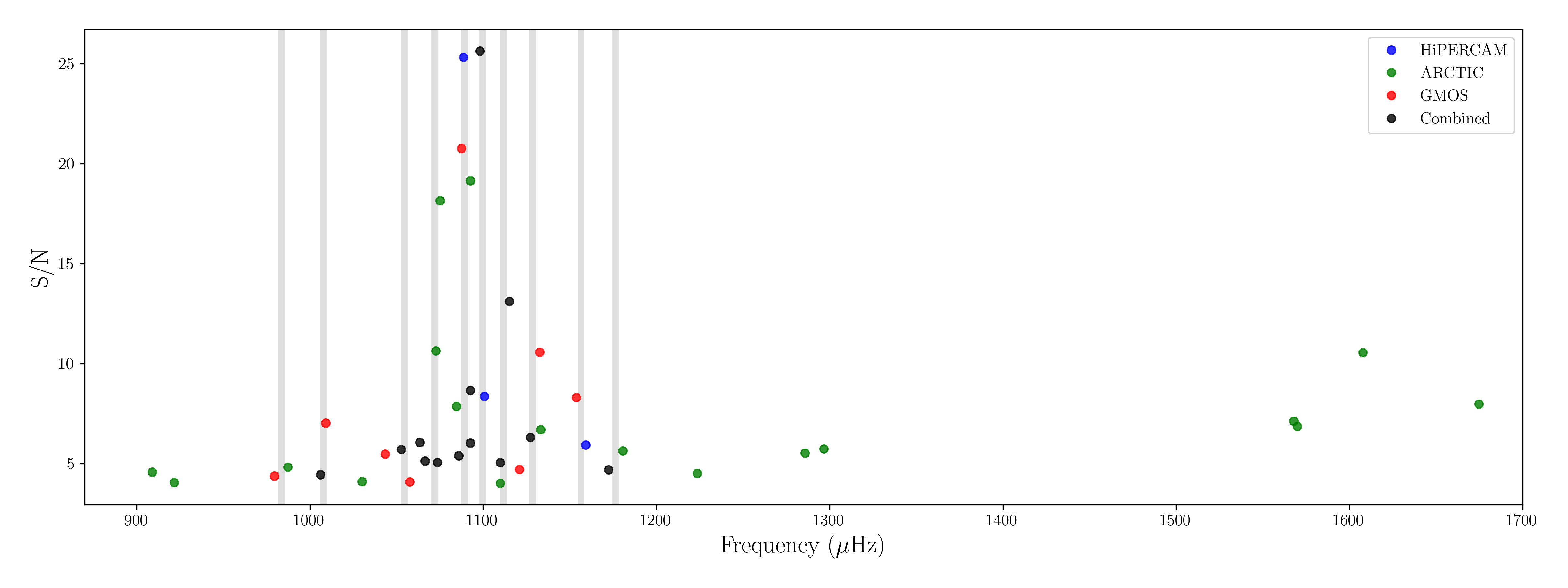} 
\caption{Detected pulsation frequencies from HiPERCAM (blue), ARCTIC (green), GMOS (red), and the combined dataset (black), plotted as a function of their signal-to-noise ratios. Each symbol marks a prewhitened frequency, with the symbol size reflecting its associated uncertainty. Grey vertical bands indicate frequencies detected in at least two independent datasets, merged within a tolerance of $\pm11~\mu$Hz, consistent with the alias width expected from the temporal sampling of our ground-based observations.}
\label{fig:F_plot}
\end{figure*}

\begin{table} 
\centering 
\scriptsize  
\setlength{\tabcolsep}{2pt} 
\caption{Frequencies, periods, and amplitudes detected from different datasets.} 
\begin{tabular}{ccccc} 
\hline 
\multicolumn{5}{c}{\textbf{Combined}}\\ 
\hline 
ID & Frequency  & Period & Amplitude & S/N \\ 
& ($\mu$Hz)  & (s) & (mma) & \\ 
\hline
 $f_{\rm 1}$  & $1006.0656 \pm 0.0040$ & $994.0239 \pm 0.0040$ & $0.1352 \pm 0.0670$ & 4.44 \\              
 $f_{\rm 2}$  & $1052.6831 \pm 0.0032$ & $950.0321 \pm 0.0029$ & $0.1842 \pm 0.0712$ & 5.70 \\              
 $f_{\rm 3}$  & $1063.5058 \pm 0.0031$ & $940.0121 \pm 0.0027$ & $0.1933 \pm 0.0703$ & 6.05 \\            
 $f_{\rm 4}$  & $1066.5095 \pm 0.0037$ & $937.9411 \pm 0.0032$ & $0.1666 \pm 0.0714$ & 5.12 \\            
 $f_{\rm 5}$  & $1073.5213 \pm 0.0037$ & $931.1222 \pm 0.0032$ & $0.1573 \pm 0.0682$ & 5.06 \\            
 $f_{\rm 6}$  & $1085.9444 \pm 0.0035$ & $921.1766 \pm 0.0030$ & $0.1734 \pm 0.0704$ & 5.38 \\             
 $f_{\rm 7}$  & $1092.5732 \pm 0.0032$ & $915.3726 \pm 0.0027$ & $0.1915 \pm 0.0697$ & 6.02 \\              
 $f_{\rm 8}$  & $1092.5870 \pm 0.0022$ & $915.3616 \pm 0.0018$ & $0.2840 \pm 0.0717$ & 8.66 \\             
 $f_{\rm 9}$  & $1098.0786 \pm 0.0008$ & $910.9657 \pm 0.0006$ & $1.0724 \pm 0.0915$ & 25.63 \\             
 $f_{\rm 10}$ & $1109.8230 \pm 0.0039$ & $901.0178 \pm 0.0031$ & $0.1563 \pm 0.0682$ & 5.05 \\              
 $f_{\rm 11}$ & $1115.0096 \pm 0.0015$ & $896.7570 \pm 0.0012$ & $0.4615 \pm 0.0775$ & 13.12 \\             
 $f_{\rm 12}$ & $1127.1139 \pm 0.0031$ & $887.4616 \pm 0.0025$ & $0.2012 \pm 0.0704$ & 6.30 \\              
 $f_{\rm 13}$ & $1172.3030 \pm 0.0044$ & $853.2857 \pm 0.0032$ & $0.1416 \pm 0.0663$ & 4.69 \\              

\hline 
\multicolumn{5}{c}{\textbf{Combination frequency}}\\ 
\hline 
$f_{\rm 7}\times2$ & $2185.6156 \pm 0.0088$ & $457.6311 \pm 0.0018$ & $0.1379 \pm 0.0696$ & 4.35 \\
\hline 

\multicolumn{5}{c}{\textbf{ARCTIC}}\\ 
\hline 
$f_{\rm 1}$& $908.9118   \pm 0.0231$ &$1100.8721 \pm 0.0280$ & $ 0.5258	\pm  0.3704 $ & 4.57 \\
$f_{\rm 2}$& $921.6559   \pm 0.0265$ &$1085.3019 \pm 0.0312$ & $ 0.3954	\pm  0.3161 $ & 4.04 \\
$f_{\rm 3}$& $987.1833   \pm 0.0238$ &$1013.0482 \pm 0.0244$ & $ 0.5590	\pm  0.3746 $ & 4.82 \\
$f_{\rm 4}$& $1030.0037  \pm 0.0292$ & $970.8737 \pm 0.0275$ & $ 0.4423	\pm  0.3487 $ & 4.10 \\
$f_{\rm 5}$& $1072.6316  \pm 0.0117$ & $932.5060 \pm 0.0102$ & $ 1.3849	\pm  0.4217 $ & 10.63 \\
$f_{\rm 6}$& $1075.1172  \pm 0.0069$ & $930.3300 \pm 0.0060$ & $ 2.3976	\pm  0.4269 $ & 18.15 \\
$f_{\rm 7}$& $1084.5961  \pm 0.0160$ & $921.9937 \pm 0.0136$ & $ 1.0114	\pm  0.4147 $ & 7.86 \\
$f_{\rm 8}$& $1092.5732  \pm 0.0066$ & $915.1831 \pm 0.0055$ & $ 2.5832	\pm  0.4362 $ & 19.15 \\
$f_{\rm 9}$& $1109.7358  \pm 0.0322$ & $901.2765 \pm 0.0261$ & $ 0.4349	\pm  0.3503 $ & 4.01 \\
$f_{\rm 10}$& $1133.1421 \pm 0.0197$ & $882.9885 \pm 0.0153$ & $ 0.8379	\pm  0.4050 $ & 6.69 \\
$f_{\rm 11}$& $1180.5163 \pm 0.0244$ & $847.2234 \pm 0.0175$ & $ 0.6803	\pm  0.3903 $ & 5.63 \\
$f_{\rm 12}$& $1223.4675 \pm 0.0316$ & $817.5270 \pm 0.0211$ & $ 0.4970	\pm  0.3561 $ & 4.50 \\
$f_{\rm 13}$& $1285.7756 \pm 0.0271$ & $778.0007 \pm 0.0164$ & $ 0.6493	\pm  0.3800 $ & 5.52 \\
$f_{\rm 14}$& $1296.6051 \pm 0.0263$ & $771.3856 \pm 0.0156$ & $ 0.6878	\pm  0.3874 $ & 5.73 \\
$f_{\rm 15}$& $1567.7327 \pm 0.0256$ & $638.0536 \pm 0.0104$ & $ 0.9035	\pm  0.4109 $ & 7.11 \\
$f_{\rm 16}$& $1569.8101 \pm 0.0266$ & $637.1223 \pm 0.0108$ & $ 0.8457	\pm  0.3982 $ & 6.86 \\
$f_{\rm 17}$& $1607.7351 \pm 0.0177$ & $621.8074 \pm 0.0068$ & $ 1.3651	\pm  0.4176 $ & 10.55 \\
$f_{\rm 18}$& $1674.7116 \pm 0.0245$ & $597.2432 \pm 0.0087$ & $ 1.0303	\pm  0.4182 $ & 7.96 \\
\hline 
\multicolumn{5}{c}{\textbf{Combination frequency}}\\ 
\hline 
$f_{\rm 5}\times2$ & $2139.5185 \pm 0.0615$ & $467.5145 \pm 0.0134$ & $ 0.4295	\pm  0.3422 $ & 4.05 \\
\hline

\multicolumn{5}{c}{\textbf{GMOS}}\\ 
\hline 
$f_{\rm 1}$&  $979.5094 \pm 0.1363$ & $1020.9035 \pm 0.1421$ & $0.5687 \pm	1.2294$ & 4.38 \\
$f_{\rm 2}$&  $1009.2020 \pm 0.0876$ & $990.8786 \pm 0.0861$ & $1.0147 \pm	1.3608$ & 7.02 \\
$f_{\rm 3}$&  $1043.4338 \pm 0.1161$ & $958.3682 \pm 0.1068$ & $0.7714 \pm	1.3324$ & 5.47 \\
$f_{\rm 4}$&  $1057.5366 \pm 0.1582$ & $945.5827 \pm 0.1413$ & $0.5162 \pm	1.2025$ & 4.07 \\   
$f_{\rm 5}$&  $1087.5735 \pm 0.0319$ & $919.5123 \pm 0.0270$ & $3.3272 \pm	1.5152$ & 20.76 \\  
$f_{\rm 6}$& $1120.8505 \pm 0.1453$ & $892.5887 \pm 0.1156$ & $0.6474 \pm	1.3066$ & 4.70 \\   
$f_{\rm 7}$& $1132.6054 \pm 0.0653$ & $882.9865 \pm 0.0509$ & $1.6196 \pm	1.4490$ & 10.56 \\  
$f_{\rm 8}$& $1153.6736 \pm 0.0847$ & $866.9812 \pm 0.0635$ & $1.2337 \pm	1.4036$ & 8.30 \\    
\hline 
\multicolumn{5}{c}{\textbf{Combination frequency}}\\ 
\hline 
$f_{\rm 5}\times2$ & $2174.6636 \pm 0.2853$ & $459.9000 \pm 0.0604$ & $0.6289 \pm	1.2832$ & 4.64 \\ 
\hline

\multicolumn{5}{c}{\textbf{HiPERCAM}}\\ 
\hline 
ID & Frequency  & Period & Amplitude & S/N \\ 
& ($\mu$Hz)  & (s) & (mma) & \\ 
\hline 
$f_{\rm 1}$ & $1088.6968 \pm 0.0009$ & $918.9061 \pm 0.0008$ & $1.0819 \pm 0.1103$ & 25.32 \\   
$f_{\rm 2}$ & $1100.6776 \pm 0.0028$ & $908.9922 \pm 0.0023$ & $0.2918 \pm 0.0902$ & 8.36 \\    
$f_{\rm 3}$ & $1159.1762 \pm 0.0042$ & $862.7262 \pm 0.0031$ & $0.2043 \pm 0.0894$ & 5.93 \\    
\hline 
\multicolumn{5}{c}{\textbf{Combination frequency}}\\ 
\hline 
$f_{\rm 1}\times2$  & $2182.1500 \pm 0.0112$ & $458.3255 \pm 0.0023$ & $0.1336 \pm 0.0828$ & 4.17 \\ 
\hline

\end{tabular} 
\label{tab:Frequencies} 
\end{table}


To identify pulsation frequencies consistently detected across multiple datasets, we applied a cross-matching criterion that accounts for both the formal frequency uncertainties and the alias structure inherent to each observation.  
Given the distinct temporal sampling of the HiPERCAM, ARCTIC, GMOS, and combined datasets, we adopted a tolerance that combines statistical and aliasing contributions.  
Two frequencies, $f_i$ and $f_j$, with respective uncertainties $\sigma_i$ and $\sigma_j$, were considered to represent the same intrinsic mode if $| f_i - f_j | \le \sqrt{(\sigma_i^2 + \sigma_j^2)} + \Delta f_{\mathrm{alias}}$, where $\Delta f_{\mathrm{alias}} = 11.57~\mu\mathrm{Hz}$ approximates the characteristic alias width of our ground-based data. This approach ensures that genuine overlapping detections are not excluded due to minor frequency shifts caused by window-function asymmetries or unresolved aliasing.  When multiple detections satisfied this condition, we retained the measurement with the highest S/N as the representative value of the mode, ensuring that the most significant determination was adopted.  
Frequencies fulfilling these criteria in two or more datasets were classified as recurrent peaks, representing the most probable pulsation modes. 
The distribution of all detected frequencies and their signal-to-noise ratios is illustrated in Figure~\ref{fig:F_plot}, while the final list of recurrent frequencies is provided in Table~\ref{tab:fl}.

\begin{table} 
\centering 
\scriptsize  
\setlength{\tabcolsep}{2pt} 
\caption{Recurrent frequencies detected in at least two independent datasets. Quoted uncertainties correspond to the mean formal errors of the merged detections. The S/N column lists the highest signal-to-noise ratio among the contributing datasets, selecting the peak with the highest S/N within $\pm 11~\mu$Hz.}
\begin{tabular}{cccc} 
\hline
Frequency & Uncertainty & S/N & Detected in \\
($\mu$Hz) & ($\mu$Hz) &  & \\
\hline
987.1833  & $\pm$0.0238 & 4.82  & ARCTIC, GMOS \\
1009.2020 & $\pm$0.0876 & 7.02  & Combined, GMOS \\
1052.6831 & $\pm$0.0032 & 5.70  & Combined, GMOS \\
1072.6316 & $\pm$0.0117 & 10.63 & ARCTIC, Combined \\
1088.6968 & $\pm$0.0009 & 25.32 & Combined, ARCTIC, GMOS, HiPERCAM \\
1098.0786 & $\pm$0.0008 & 25.63 & Combined, HiPERCAM \\
1109.8230 & $\pm$0.0039 & 5.05  & ARCTIC, Combined \\
1127.1139 & $\pm$0.0031 & 6.30  & ARCTIC, Combined, GMOS \\
1153.6736 & $\pm$0.0847 & 8.30  & GMOS, HiPERCAM \\
1180.5163 & $\pm$0.0244 & 5.63  & ARCTIC, Combined, \cite{2020NatAs...4..663H} \\
\hline
\end{tabular}
\label{tab:fl}
\end{table}

This analysis reveals ten recurrent frequencies that are consistently detected across two or more independent datasets. These represent the most significant and probable candidates for intrinsic pulsation modes of J0551. However, given the limited temporal coverage and possible amplitude variability between datasets, we cannot yet confirm that all of these correspond to genuine stellar pulsations rather than manifestations of amplitude modulation or window-function effects.

\section{The mass of WD~J0551+4135}
\label{evolutionary_models}

\citet{kilic24} obtained $T_{\rm eff}= 12\,997 \pm 115$ K and $R=0.00630^{+0.00005}_{-0.00006}~R_\odot$ for J0551+4135,
which imply $\log{g} = 8.90 \pm 0.01$ and $M=1.139 \pm 0.005~M_\odot$ based on the evolutionary models from \citet{Bedard2020} with CO cores,
$q({\rm He})\equiv \log M_{\rm   He}/M_{\star}=10^{-2}$ and $q({\rm H})=10^{-10}$, which are representative of thin H-atmosphere
white dwarfs. These values are consistent with $T_{\rm eff} = 13\,370 \pm 330$ K and $\log{g} = 8.908 \pm 0.014$ from \citet{2020NatAs...4..663H}. 

We employed a new grid of models for ultra-massive WDs with H- and C-rich atmospheres representative of DAQ WDs computed using {\tt LPCODE} stellar evolutionary code, developed by the La Plata group \citep{2005A&A...435..631A,2015A&A...576A...9A,2016A&A...588A..25M}, assuming both CO and ONe-core chemical compositions.

\begin{figure}[htp]
   \centering
   \includegraphics[width=1\columnwidth]{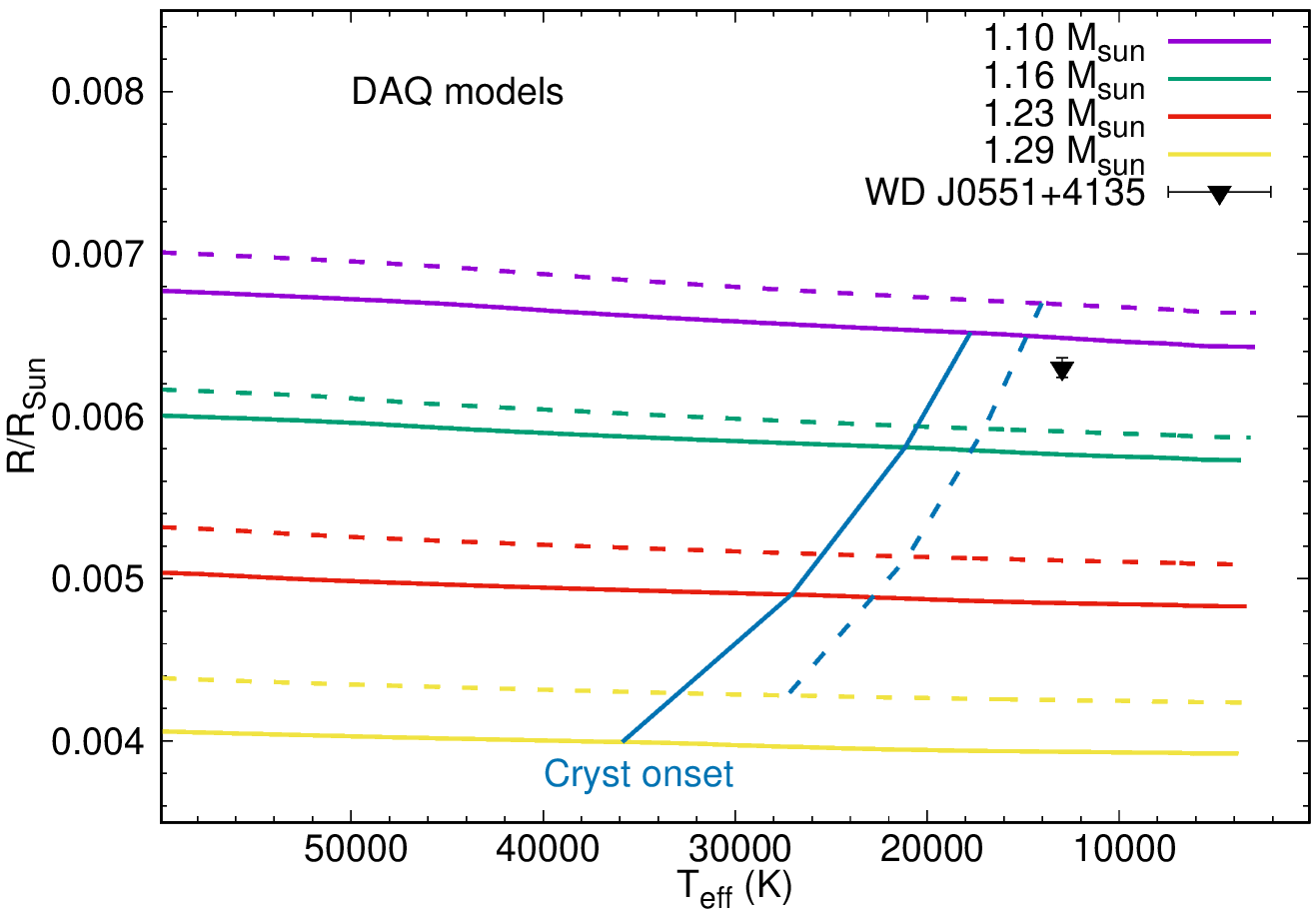} 
 \caption{Evolutionary tracks in the Radius-effective temperature plane for our DAQ models with different stellar masses and two core compositions. Solid (dashed) lines refer to ONe (CO) core composition. The diagonal solid and dashed lines mark the onset of crystallization.}
    \label{fig:logg}
\end{figure}

Figure \ref{fig:logg} shows the evolutionary tracks for our DAQ models with masses ranging from 1.10 to $1.29~M_\odot$.
Solid and dashed lines show the ONe and CO core models, respectively. The diagonal solid and dashed lines mark the onset
of crystallization for reference. These models have total He- and H-masses of $10^{-10} M_\odot$. Rapid
gravitational settling leaves a pure H envelope at the early stages of the WD phase. However, chemical
diffusion erases the He buffer, thus leaving a H/C interphase \citep[see Figure 3 of][]{2020NatAs...4..663H}. Once
the outer convective zone reaches the H-C interphase, C is dredged-up to the surface in a H-dominated atmosphere. This results in a spectrum dominated by H, with
noticeable traces of C and no detectable He. \citet{2020NatAs...4..663H} employed similar models, but here, we
also consider the effects of gravitational settling of ions due to Coulomb interactions at high densities
\citep{2020A&A...644A..55A} and employ a larger grid of models to cover the range of
masses expected for ultra-massive WDs. Interpolating in our model grid for the effective temperature and radius of J0551+4135, we obtain a mass of $1.13 \pm 0.01M_\odot$ and a cooling age of $1.7 \pm 0.1$ Gyr for a
CO core white dwarf, and $1.12 \pm 0.01 M_\odot$ and $1.6 \pm 0.1$ Gyr for an ONe core object, respectively. Note that
these cooling ages are likely underestimated, since J0551+4135 is kinematically old \citep{kilic24}.  

Even though these models are appropriate for describing the spectroscopic characteristics of DAQ white dwarfs,
they are not suitable for seismological analyses since they are not based on realistic evolutionary scenarios for
the formation of DAQ white dwarfs, and they have an almost arbitrary internal chemical structure. Calculations of those
evolutionary models are beyond the scope of this work, and we leave detailed asteroseismology of this unique pulsating
DAQ white dwarf to future work.

\section{Conclusions and future prospects}
\label{conclusions}

We report the discovery of a remarkably rich pulsation spectrum in the ultra-massive DAQ white dwarf J0551+4135.
 Using multi-site time-series photometry from APO, Gemini, and GTC/HiPERCAM over 19 nights, we detected over 10 potential pulsational frequencies. This object stands out not only for its atmospheric composition —hydrogen-dominated with substantial carbon enrichment— but also for its high mass and space velocity, which favor a merger origin.

Model fits to the spectral energy distribution of J0551+4135 along with the Gaia parallax indicate $T_{\rm eff}\approx13\,000$ K and
$R=0.0063~R_\odot$ \citep{2020NatAs...4..663H,kilic24}. To interpret these parameters, we used an extended model grid of ultra-massive WD
evolutionary models computed with the {\tt LPCODE} by the La Plata group, considering both CO- and ONe-core compositions. These models, while not suitable for seismological analysis, can be used to characterize the spectroscopic properties of DAQ white dwarfs. By interpolating within this grid using the temperature and radius constraints, we obtain a mass of $1.13 \pm 0.01,M_\odot$ and a cooling age of $1.7 \pm 0.2$ Gyr for CO-core models, and $1.12 \pm 0.01,M_\odot$ and $1.6 \pm 0.1$ Gyr for ONe-core models, respectively. 

The pulsation spectrum exhibits a complex pattern of closely spaced and potentially overlapping modes, complicating the identification of period spacings and rotational splittings. Nevertheless, the stable detection of several modes across multiple campaigns provides a robust basis for future mode identification. 

Given its unique classification as a DAQ white dwarf and potential connection to merger evolution scenarios, the final list of significant frequencies presented in this paper (see Table \ref{tab:Frequencies}) renders this star an exceptional target for future asteroseismic studies. These studies will aim to probe the star's internal structure, rotational properties, and evolutionary origin. Future research will focus on generating a new set of fully evolutionary models for DAQ white dwarfs, specifically designed for asteroseismology. 
In addition, further observations —ideally multi-site campaigns to reduce daily aliasing and to improve the S/N ratio— will be crucial for securely identifying the pulsation modes and constraining the star’s internal structure.


\begin{acknowledgments}

The authors thank S. O. Kepler for his detailed suggestions on the frequency analysis and for his valuable comments and feedback that improved the quality of this paper. 
M. U. gratefully acknowledges funding from the Research Foundation Flanders (FWO) by means of a junior postdoctoral fellowship (grant agreement No. 1247624N). 
This work is supported in part by the NSF under grant  AST-2508429, and the NASA
under grants 80NSSC22K0479, 80NSSC24K0380, and 80NSSC24K0436.
This work was supported by PIP 112-200801-00940 grant from CONICET, grant G149 from the University of La Plata, PIP-2971 from CONICET (Argentina) and by PICT 2020-03316 from Agencia I+D+i (Argentina). 
This work was partially supported by the MINECO grant  PID2023-148661NB-I00 and by the AGAUR/Generalitat de Catalunya grant SGR-386/2021. M. C. acknowledges grant RYC2021-032721-I, funded by
MCIN/AEI/10.13039/501100011033 and by the European
Union NextGenerationEU/PRTR.

Based on observations made with the Gran Telescopio Canarias  (Prog. ID: GTC19-24B), installed at the Spanish Observatorio del Roque de los Muchachos of the Instituto de Astrofísica de Canarias, on the island of La Palma. 

The Apache Point Observatory 3.5-meter telescope is owned and operated by the Astrophysical Research Consortium. 

Based on observations obtained at the international Gemini Observatory, a program of NSF's NOIRLab, which is managed by the Association of Universities for Research in Astronomy (AURA) under a cooperative agreement with the National Science Foundation on behalf of the Gemini Observatory partnership: the National Science Foundation (United States), National Research Council (Canada), Agencia Nacional de Investigaci\'{o}n y Desarrollo (Chile), Ministerio de Ciencia, Tecnolog\'{i}a e Innovaci\'{o}n (Argentina), Minist\'{e}rio da Ci\^{e}ncia, Tecnologia, Inova\c{c}\~{o}es e Comunica\c{c}\~{o}es (Brazil), and Korea Astronomy and Space Science Institute (Republic of Korea). This research has made use of the NASA Astrophysics Data System.


\end{acknowledgments}

%

\vspace{5mm}
\facilities{GTC (HiPERCAM), ARC 3.5m (ARCTIC), Gemini:Gillett (GMOS spectrograph)}




\bibliography{myref}{}
\bibliographystyle{aasjournal}

\appendix
\section{Light curves and frequency solution for individual dataset}\label{LC}

\begin{figure*}
   \includegraphics[width=1.0\textwidth]{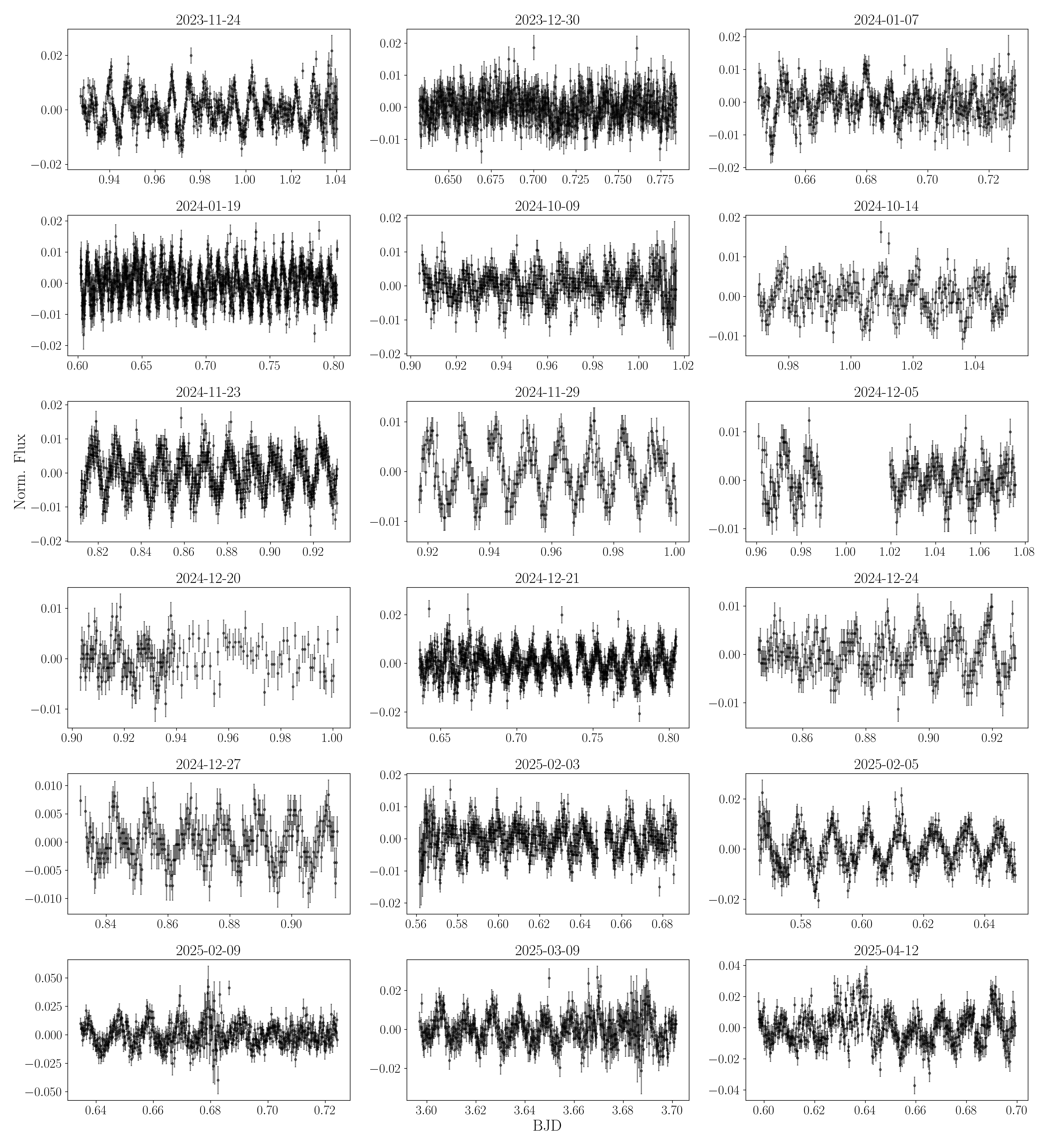} 
 \caption{Light curves obtained with APO and Gemini. }
    \label{fig:LC_APO}
\end{figure*}

\begin{figure*}
   \includegraphics[width=1.0\textwidth]{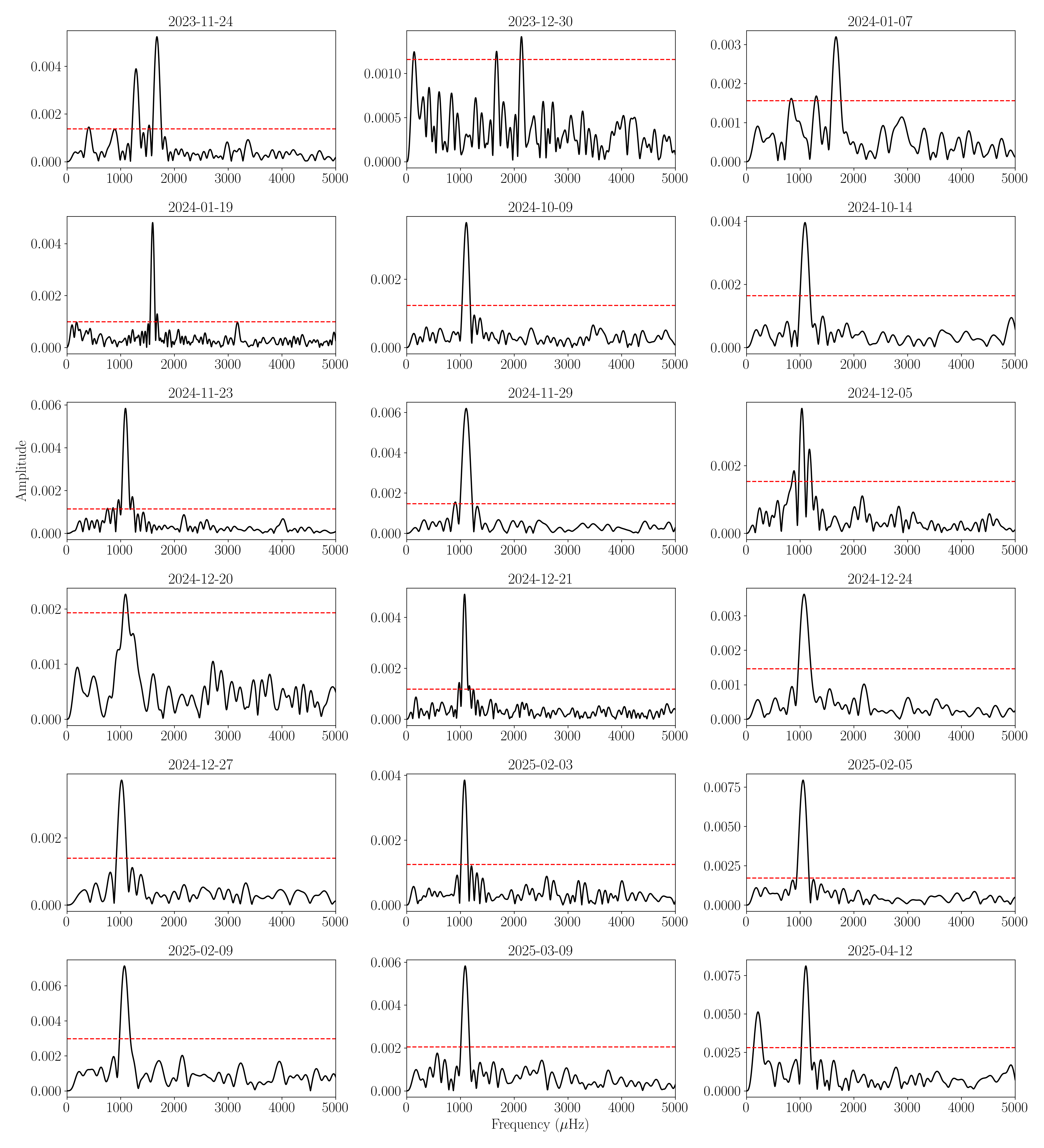} 
 \caption{Fourier transforms of the light curves obtained with APO and Gemini. }
    \label{fig:FT_APO}
\end{figure*}

\begin{figure*}[ht]
\centering
    \includegraphics[width=0.49\textwidth]{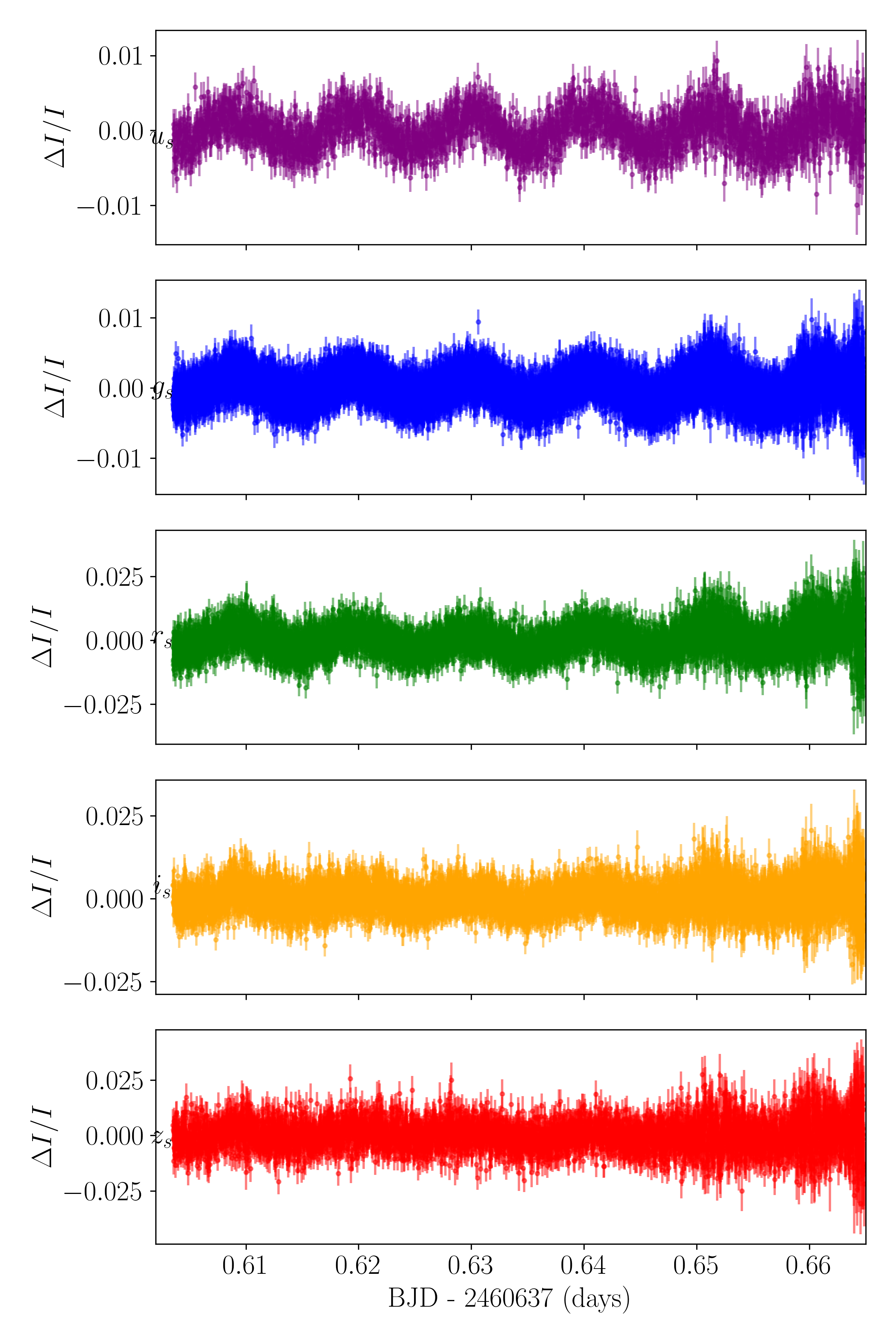}
    \label{fig:LC_GTC}
\hfill
    \includegraphics[width=0.49\textwidth]{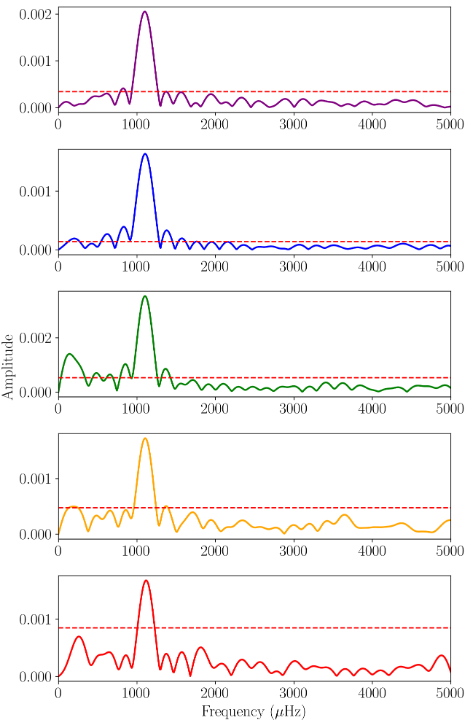}
    \label{fig:FT_GTC}

\caption{Left: Light curve of the target obtained with HiPERCAM on 2024-11-22 (see Section~\ref{analysis}). HiPERCAM filters $u_{s},g_{s},r_{s},i_{s},z_{s}$ are shown from top to bottom. Right: Corresponding Fourier transform of the same light curve, illustrating the detected periodicities.}
\label{fig:GTC_LC_and_FT}
\end{figure*}

\begin{figure*}[ht]
\centering
    \includegraphics[width=0.49\textwidth]{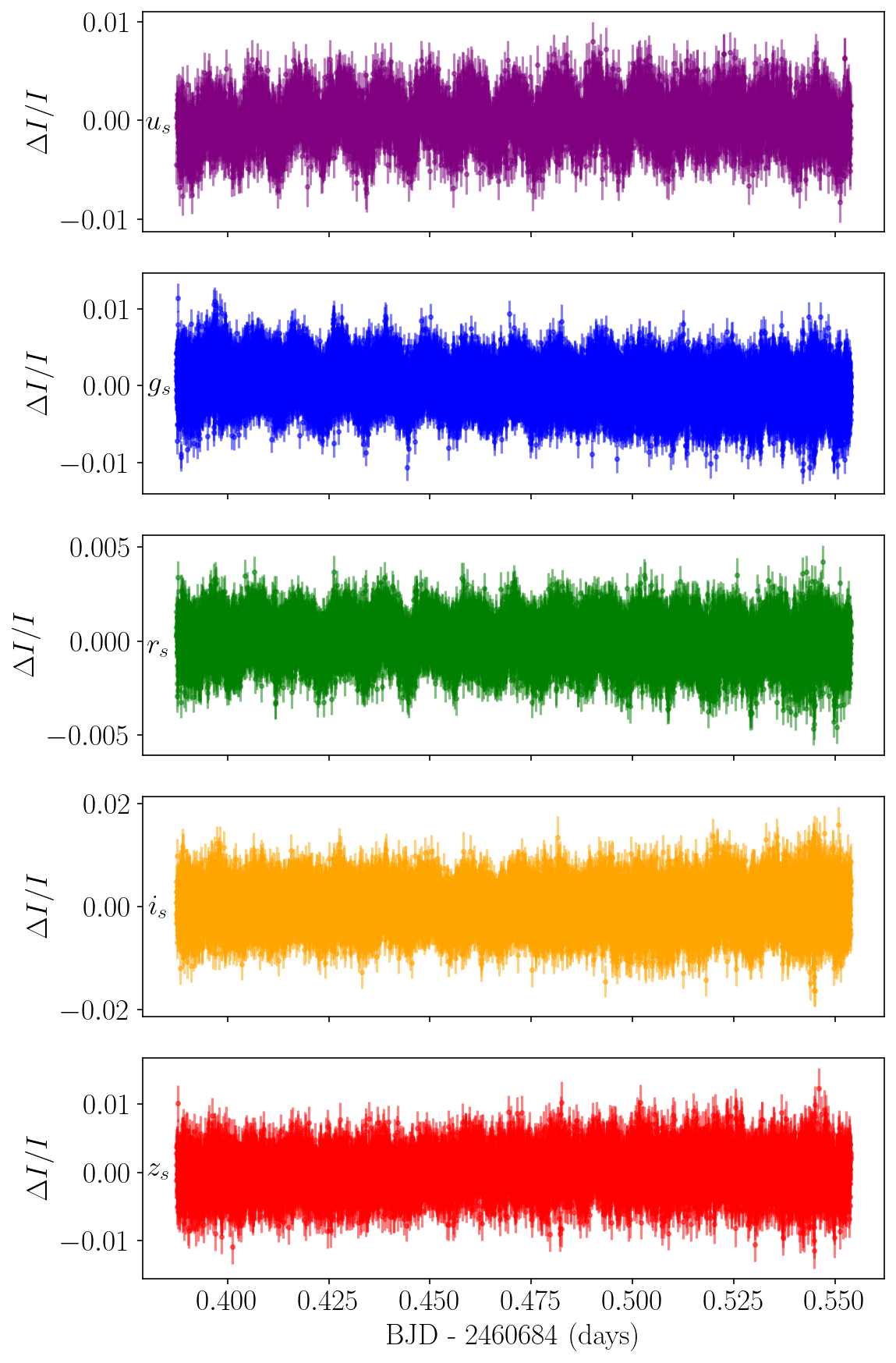}
    \label{fig:LC_GTC}
\hfill
    \includegraphics[width=0.49\textwidth]{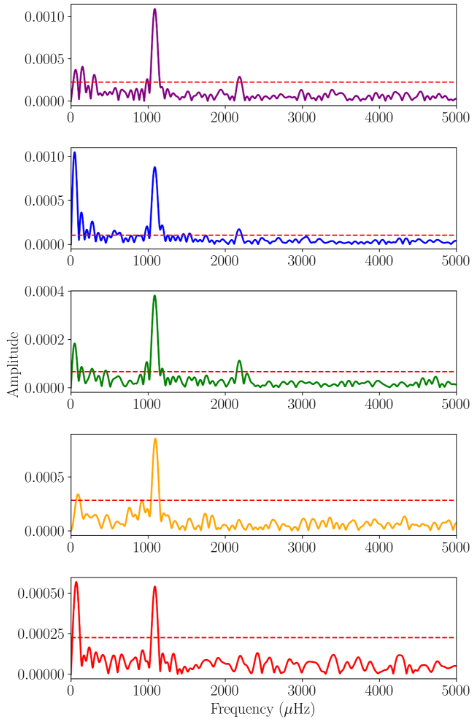}
    \label{fig:FT_GTC}

\caption{Left: Light curve of the target obtained with HiPERCAM on 2025-01-08 (see Section~\ref{analysis}). HiPERCAM filters $u_{s},g_{s},r_{s},i_{s},z_{s}$ are shown from top to bottom. Right: Corresponding Fourier transform of the same light curve, illustrating the detected periodicities.}
\label{fig:GTC_LC_and_FT_2}
\end{figure*}

We obtained time-series photometry of J0551 with APO/ARCTIC, Gemini/GMOS, and GTC/HiPERCAM. 
Figures~\ref{fig:LC_APO} and \ref{fig:FT_APO} show the light curves and corresponding Fourier transforms from the APO and Gemini runs. 
HiPERCAM observations were conducted on two separate nights, 2024-11-22 and 2025-01-08. 
The resulting light curves in the five HiPERCAM filters ($u_{s}, g_{s}, r_{s}, i_{s}, z_{s}$) and their corresponding Fourier transforms are presented in Figure~\ref{fig:GTC_LC_and_FT} (first night) and \ref{fig:GTC_LC_and_FT_2} (second night). 
These data form the basis for the frequency solution of J0551+413, which is presented in Table~\ref{table:Full_F_list}.

\begin{table*}[ht]
\setlength{\tabcolsep}{8pt}
\renewcommand{\arraystretch}{.9}
\centering
\caption{Frequency solution for each night for J0551.}
\label{table:Full_F_list}
\begin{tabular}{ccccr}
\hline
\noalign{\smallskip}
$\nu$ & $\Pi$ & $A$ & S/N \\
 ($\mu$Hz) & (s) &  & & \\
\noalign{\smallskip}
\hline
\noalign{\smallskip}


\multicolumn{5}{c}{\textbf{ARCTIC (2023-11-24)}} \\

$1281.1373 \pm 3.0856$ & $780.5565 \pm 1.8800$ & $0.0037 \pm 0.0002$ & $12.08$ \\
$1678.4510 \pm 2.1900$ & $595.7874 \pm 0.7774$ & $0.0052 \pm 0.0002$ & $17.02$ \\

\noalign{\smallskip}
\hline
\noalign{\smallskip}

\multicolumn{5}{c}{\textbf{ARCTIC (2023-12-30)}} \\

$1679.5322 \pm 6.8227$ & $595.4039 \pm 2.4187$ & $0.0013 \pm 0.0002$ & $4.87$ \\
$2134.8040 \pm 6.0271$ & $468.4271 \pm 1.3225$ & $0.0014 \pm 0.0002$ & $5.55$ \\

\noalign{\smallskip}
\hline
\noalign{\smallskip}

\multicolumn{5}{c}{\textbf{ARCTIC (2024-01-07)}} \\

$1668.2957 \pm 6.6064$ & $599.4141 \pm 2.3737$ & $0.0032 \pm 0.0003$ & $9.16$ \\

\noalign{\smallskip}
\hline
\noalign{\smallskip}

\multicolumn{5}{c}{\textbf{ARCTIC (2024-01-19)}} \\

$1595.4570 \pm 1.0800$ & $626.7797 \pm 0.4243$ & $0.0048 \pm 0.0002$ & $21.94$ \\
$3176.0362 \pm 5.3308$ & $314.8579 \pm 0.5285$ & $0.0010 \pm 0.0002$ & $4.48$ \\

\noalign{\smallskip}
\hline
\noalign{\smallskip}

\multicolumn{5}{c}{\textbf{ARCTIC (2024-10-09)}} \\
$1110.5566 \pm 0.0429$ & $900.6450 \pm 34.7600$ & $0.2682 \pm 0.0002$ & $9.93$ \\

\noalign{\smallskip}
\hline
\noalign{\smallskip}

\multicolumn{5}{c}{\textbf{GMOS (2024-10-14)}} \\
$1091.5378 \pm 5.1518$ & $916.1387 \pm 4.3240$ & $0.0039 \pm 0.0003$ & $10.78$ \\

\noalign{\smallskip}
\hline
\noalign{\smallskip}

\multicolumn{5}{c}{\textbf{ARCTIC (2024-11-23)}} \\
$1091.4791 \pm 1.6997$ & $916.1880 \pm 1.4267$ & $0.0058 \pm 0.0002$ & $22.74$ \\
$1193.7252 \pm 8.5344$ & $837.7137 \pm 5.9891$ & $0.0011 \pm 0.0002$ & $4.53$ \\

\noalign{\smallskip}
\hline
\noalign{\smallskip}

\multicolumn{5}{c}{\textbf{GMOS (2024-11-29 )}} \\
$1108.0355 \pm 2.6042$ & $902.4982 \pm 2.1211$ & $0.0062 \pm 0.0002$ & $19.03$ \\

\noalign{\smallskip}
\hline
\noalign{\smallskip}

\multicolumn{5}{c}{\textbf{GMOS (2024-12-05 )}} \\
$1034.6252 \pm 3.54$ & $966.5336 \pm 3.307$ & $0.0037 \pm 0.0002$ & $10.98$ \\

\noalign{\smallskip}
\hline
\noalign{\smallskip}

\multicolumn{5}{c}{\textbf{GMOS (2024-12-20  )}} \\
$1087.7868 \pm 9.1577$ & $919.2978 \pm 7.7392$ & $0.0023 \pm 0.0003$ & $5.3$ \\

\noalign{\smallskip}
\hline
\noalign{\smallskip}

\multicolumn{5}{c}{\textbf{ARCTIC (2024-12-21)}} \\
$1078.7351 \pm 1.5167$ & $927.0116 \pm 1.3033$ & $0.0049 \pm 0.0002$ & $18.82$ \\
$1139.2991 \pm 6.1623$ & $877.7326 \pm 4.7475$ & $0.0012 \pm 0.0002$ & $4.63$ \\

\noalign{\smallskip}
\hline
\noalign{\smallskip}

\multicolumn{5}{c}{\textbf{GMOS (2024-12-24)}} \\
$1018.7525 \pm 4.3544$ & $981.5927 \pm 4.1956$ & $0.0038 \pm 0.0002$ & $12.22$ \\

\noalign{\smallskip}
\hline
\noalign{\smallskip}

\multicolumn{5}{c}{\textbf{GMOS (2024-12-27)}} \\
$1055.7849 \pm 5.1529$ & $947.1626 \pm 4.6227$ & $0.0033 \pm 0.0002$ & $10.25$ \\

\noalign{\smallskip}
\hline

\end{tabular}
\end{table*}
\par\medskip

\begin{table*}[ht]
\renewcommand\thetable{2}
\renewcommand{\arraystretch}{.9}

\caption{Full frequency solution for J0551+4135 (continued).}
\begin{tabular}{ccccr}
\hline
\noalign{\smallskip}
$\nu$ & $\Pi$ & $A$ & S/N \\
 ($\mu$Hz) & (s) & & & \\
\noalign{\smallskip}
\hline
\noalign{\smallskip}

\multicolumn{5}{c}{\textbf{ARCTIC (2025-02-03)}} \\
$1054.0597 \pm 2.5281$ & $948.7128 \pm 2.2754$ & $0.0079 \pm 0.0003$ & $20.77$ \\

\noalign{\smallskip}
\hline
\noalign{\smallskip}

\multicolumn{5}{c}{\textbf{ARCTIC (2025-02-05)}} \\
$1061.3919 \pm 9.5938$ & $942.1591 \pm 8.5160$ & $0.0067 \pm 0.0007$ & $10.15$ \\

\noalign{\smallskip}
\hline
\noalign{\smallskip}

\multicolumn{5}{c}{\textbf{ARCTIC (2025-02-09)}} \\
$1077.7814 \pm 2.7571$ & $927.8320 \pm 2.3735$ & $0.0038 \pm 0.0002$ & $13.75$ \\
$1089.5544 \pm 3.6582$ & $917.8064 \pm 3.0816$ & $0.0058 \pm 0.0003$ & $12.73$ \\

\noalign{\smallskip}
\hline
\noalign{\smallskip}

\multicolumn{5}{c}{\textbf{ARCTIC (2025-03-09)}} \\
$1106.4487 \pm 3.8070$ & $903.7925 \pm 3.1097$ & $0.0082 \pm 0.0005$ & $13.18$ \\

\noalign{\smallskip}
\hline
\noalign{\smallskip}

\multicolumn{5}{c}{\textbf{ARCTIC (2025-04-12)}} \\
$1146.4526 \pm 18.8728$ & $872.2559 \pm 14.3590$ & $0.0033 \pm 0.0006$ & $5.06$ \\

\noalign{\smallskip}
\hline
\noalign{\smallskip}

\multicolumn{5}{c}{\textbf{HiPERCAM (2024-11-22)}} \\
\noalign{\smallskip}
\hline
\noalign{\smallskip}
\multicolumn{5}{c}{\textbf{$u_{s}$}} \\
$1104.3647 \pm 2.9494$ & $905.4707 \pm 2.4173$ & $0.0021 \pm 0.0001$ & 27.12 \\

\multicolumn{5}{c}{\textbf{$g_{s}$}} \\
$1103.8914 \pm 1.5521$ & $905.8442 \pm 1.2739$ & $0.0016 \pm 0.0000$ & 52.75 \\

\multicolumn{5}{c}{\textbf{$r_{s}$}} \\
$1103.9144 \pm 2.7321$ & $905.8251 \pm 2.2405$ & $0.0035 \pm 0.0001$ & 30.22 \\

\multicolumn{5}{c}{\textbf{$i_{s}$}} \\
$1103.9144 \pm 5.0052$ & $905.8251 \pm 4.1072$ & $0.0017 \pm 0.0001$ & 16.42 \\

\multicolumn{5}{c}{\textbf{$z_{s}$}} \\
$1113.0445 \pm 9.2954$ & $898.6455 \pm 7.5041$ & $0.0017 \pm 0.0002$ & 8.90 \\

\noalign{\smallskip}
\hline
\noalign{\smallskip}

\multicolumn{5}{c}{\textbf{HiPERCAM (2025-01-08)}} \\
\noalign{\smallskip}
\hline
\noalign{\smallskip}
\multicolumn{5}{c}{\textbf{$u_{s}$}} \\
$1088.8140 \pm 12.4187$ & $918.4465 \pm 10.4722$ & $0.0011 \pm 0.00005$ & 24.37 \\
$2189.8255 \pm 4.8186$ & $456.4877 \pm 1.0058$ & $0.0003 \pm 0.00009$ & 6.25 \\

\multicolumn{5}{c}{\textbf{$g_{s}$}} \\
$1088.6837 \pm 0.7211$ & $918.5585 \pm 0.6086$ & $0.0009 \pm 0.00004$ & 42.38 \\
$2186.2081 \pm 4.0256$ & $457.2514 \pm 0.8424$ & $0.0002 \pm 0.00008$ & 7.60 \\

\multicolumn{5}{c}{\textbf{$r_{s}$}} \\
$1084.8194 \pm 1.0406$ & $921.8132 \pm 0.8846$ & $0.0004 \pm 0.00004$ & 28.98 \\
$2186.2237 \pm 3.7328$ & $457.2482 \pm 0.7815$ & $0.0001 \pm 0.00008$ & 8.21 \\

\multicolumn{5}{c}{\textbf{$i_{s}$}} \\
$1095.4606 \pm 1.9671$ & $912.8300 \pm 1.6390$ & $0.0009 \pm 0.00004$ & 15.27 \\

\multicolumn{5}{c}{\textbf{$z_{s}$}} \\
$1088.7851 \pm 2.5780$ & $918.4765 \pm 2.1747$ & $0.0005 \pm 0.00005$ & 11.82 \\

\noalign{\smallskip}
\hline
\end{tabular}
\end{table*}



\end{document}